\documentstyle[epsf]{mn}

\newif\ifAMStwofonts

\def\simgt{\hbox{\rlap{\raise 0.425ex\hbox{$>$}}\lower 0.65ex\hbox{$\sim$}}}
\def\simlt{\hbox{\rlap{\raise 0.425ex\hbox{$<$}}\lower 0.65ex\hbox{$\sim$}}}

\def\PsfigVersion{1.10}
\def\setDriver{\DvipsDriver} 
\ifx\undefined\psfig\else \fi
\let\LaTeXAtSign=\@
\let\@=\relax
\edef\psfigRestoreAt{\catcode`\@=\number\catcode`@\relax}
\catcode`\@=11\relax
\newwrite\@unused
\def\ps@typeout#1{{\let\protect\string\immediate\write\@unused{#1}}}

\def\DvipsDriver{
	\ps@typeout{psfig/tex \PsfigVersion -dvips}
\def\PsfigSpecials{\DvipsSpecials} 	\def\ps@dir{/}
\def\ps@predir{} }
\def\OzTeXDriver{
	\ps@typeout{psfig/tex \PsfigVersion -oztex}
	\def\PsfigSpecials{\OzTeXSpecials}
	\def\ps@dir{:}
	\def\ps@predir{:}
	\catcode`\^^J=5
}

\def\figurepath{./:}

\def\DoPaths#1{\expandafter\EachPath#1\stoplist}
\def\leer{}
\def\EachPath#1:#2\stoplist{
  \ExistsFile{#1}{\SearchedFile}
  \ifx#2\leer
  \else
    \expandafter\EachPath#2\stoplist
  \fi}
\def\ps@dir{/}
\def\ExistsFile#1#2{%
   \openin1=\ps@predir#1\ps@dir#2
   \ifeof1
       \closein1
   \else
       \closein1
        \ifx\ps@founddir\leer
           \edef\ps@founddir{#1}
        \fi
   \fi}
\def\get@dir#1{%
  \def\ps@founddir{}
  \def\SearchedFile{#1}
  \DoPaths\figurepath
}

\def\@nnil{\@nil}
\def\@empty{}
\def\@psdonoop#1\@@#2#3{}
\def\@psdo#1:=#2\do#3{\edef\@psdotmp{#2}\ifx\@psdotmp\@empty \else
    \expandafter\@psdoloop#2,\@nil,\@nil\@@#1{#3}\fi}
\def\@psdoloop#1,#2,#3\@@#4#5{\def#4{#1}\ifx #4\@nnil \else
       #5\def#4{#2}\ifx #4\@nnil \else#5\@ipsdoloop #3\@@#4{#5}\fi\fi}
\def\@ipsdoloop#1,#2\@@#3#4{\def#3{#1}\ifx #3\@nnil 
       \let\@nextwhile=\@psdonoop \else
      #4\relax\let\@nextwhile=\@ipsdoloop\fi\@nextwhile#2\@@#3{#4}}
\def\@tpsdo#1:=#2\do#3{\xdef\@psdotmp{#2}\ifx\@psdotmp\@empty \else
    \@tpsdoloop#2\@nil\@nil\@@#1{#3}\fi}
\def\@tpsdoloop#1#2\@@#3#4{\def#3{#1}\ifx #3\@nnil 
       \let\@nextwhile=\@psdonoop \else
      #4\relax\let\@nextwhile=\@tpsdoloop\fi\@nextwhile#2\@@#3{#4}}
\ifx\undefined\fbox
\newdimen\fboxrule
\newdimen\fboxsep
\newdimen\ps@tempdima
\newbox\ps@tempboxa
\long\def\fbox#1{\leavevmode\setbox\ps@tempboxa\hbox{#1}\ps@tempdima\fboxrule
    \advance\ps@tempdima \fboxsep \advance\ps@tempdima \dp\ps@tempboxa
   \hbox{\lower \ps@tempdima\hbox
  {\vbox{\hrule height \fboxrule
          \hbox{\vrule width \fboxrule \hskip\fboxsep
          \vbox{\vskip\fboxsep \box\ps@tempboxa\vskip\fboxsep}\hskip 
                 \fboxsep\vrule width \fboxrule}
                 \hrule height \fboxrule}}}}
\fi
\newread\ps@stream
\newif\ifnot@eof       
\newif\if@noisy        
\newif\if@atend        
\newif\if@psfile       
{\catcode`\%=12\global\gdef\epsf@start{
\def\epsf@PS{PS}
\def\epsf@getbb#1{%
\openin\ps@stream=\ps@predir#1
\ifeof\ps@stream\ps@typeout{Error, File #1 not found}\else
   {\not@eoftrue \chardef\other=12
    \def\do##1{\catcode`##1=\other}\dospecials \catcode`\ =10
    \loop
       \if@psfile
	  \read\ps@stream to \epsf@fileline
       \else{
	  \obeyspaces
          \read\ps@stream to \epsf@tmp\global\let\epsf@fileline\epsf@tmp}
       \fi
       \ifeof\ps@stream\not@eoffalse\else
       \if@psfile\else
       \expandafter\epsf@test\epsf@fileline:. \\%
       \fi
          \expandafter\epsf@aux\epsf@fileline:. \\%
       \fi
   \ifnot@eof\repeat
   }\closein\ps@stream\fi}%
\long\def\epsf@test#1#2#3:#4\\{\def\epsf@testit{#1#2}
			\ifx\epsf@testit\epsf@start\else
\ps@typeout{Warning! File does not start with `\epsf@start'.  It may not be a PostScript file.}
			\fi
			\@psfiletrue} 
{\catcode`\%=12\global\let\epsf@percent=
\long\def\epsf@aux#1#2:#3\\{\ifx#1\epsf@percent
   \def\epsf@testit{#2}\ifx\epsf@testit\epsf@bblit
	\@atendfalse
        \epsf@atend #3 . \\%
	\if@atend	
	   \if@verbose{
		\ps@typeout{psfig: found `(atend)'; continuing search}
	   }\fi
        \else
        \epsf@grab #3 . . . \\%
        \not@eoffalse
        \global\no@bbfalse
        \fi
   \fi\fi}%
\def\epsf@grab #1 #2 #3 #4 #5\\{%
   \global\def\epsf@llx{#1}\ifx\epsf@llx\empty
      \epsf@grab #2 #3 #4 #5 .\\\else
   \global\def\epsf@lly{#2}%
   \global\def\epsf@urx{#3}\global\def\epsf@ury{#4}\fi}%
\def\epsf@atendlit{(atend)} 
\def\epsf@atend #1 #2 #3\\{%
   \def\epsf@tmp{#1}\ifx\epsf@tmp\empty
      \epsf@atend #2 #3 .\\\else
   \ifx\epsf@tmp\epsf@atendlit\@atendtrue\fi\fi}

\chardef\psletter = 11 
\chardef\other = 12

\newif \ifdebug 
\newif\ifc@mpute 
\c@mputetrue 

\let\then = \relax
\def\r@dian{pt }
\let\r@dians = \r@dian
\let\dimensionless@nit = \r@dian
\let\dimensionless@nits = \dimensionless@nit
\def\internal@nit{sp }
\let\internal@nits = \internal@nit
\newif\ifstillc@nverging
\def \Mess@ge #1{\ifdebug \then \message {#1} \fi}

{ 
	\catcode `\@ = \psletter
	\gdef \nodimen {\expandafter \n@dimen \the \dimen}
	\gdef \term #1 #2 #3%
	       {\edef \t@ {\the #1}
		\edef \t@@ {\expandafter \n@dimen \the #2\r@dian}%
		\t@rm {\t@} {\t@@} {#3}%
	       }
	\gdef \t@rm #1 #2 #3%
	       {{%
		\count 0 = 0
		\dimen 0 = 1 \dimensionless@nit
		\dimen 2 = #2\relax
		\Mess@ge {Calculating term #1 of \nodimen 2}%
		\loop
		\ifnum	\count 0 < #1
		\then	\advance \count 0 by 1
			\Mess@ge {Iteration \the \count 0 \space}%
			\Multiply \dimen 0 by {\dimen 2}%
			\Mess@ge {After multiplication, term = \nodimen 0}%
			\Divide \dimen 0 by {\count 0}%
			\Mess@ge {After division, term = \nodimen 0}%
		\repeat
		\Mess@ge {Final value for term #1 of 
				\nodimen 2 \space is \nodimen 0}%
		\xdef \Term {#3 = \nodimen 0 \r@dians}%
		\aftergroup \Term
	       }}
	\catcode `\p = \other
	\catcode `\t = \other
	\gdef \n@dimen #1pt{#1} 
}

\def \Divide #1by #2{\divide #1 by #2} 

\def \Multiply #1by #2
       {{
	\count 0 = #1\relax
	\count 2 = #2\relax
	\count 4 = 65536
	\Mess@ge {Before scaling, count 0 = \the \count 0 \space and
			count 2 = \the \count 2}%
	\ifnum	\count 0 > 32767 
	\then	\divide \count 0 by 4
		\divide \count 4 by 4
	\else	\ifnum	\count 0 < -32767
		\then	\divide \count 0 by 4
			\divide \count 4 by 4
		\else
		\fi
	\fi
	\ifnum	\count 2 > 32767 
	\then	\divide \count 2 by 4
		\divide \count 4 by 4
	\else	\ifnum	\count 2 < -32767
		\then	\divide \count 2 by 4
			\divide \count 4 by 4
		\else
		\fi
	\fi
	\multiply \count 0 by \count 2
	\divide \count 0 by \count 4
	\xdef \product {#1 = \the \count 0 \internal@nits}%
	\aftergroup \product
       }}

\def\r@duce{\ifdim\dimen0 > 90\r@dian \then   
		\multiply\dimen0 by -1
		\advance\dimen0 by 180\r@dian
		\r@duce
	    \else \ifdim\dimen0 < -90\r@dian \then  
		\advance\dimen0 by 360\r@dian
		\r@duce
		\fi
	    \fi}

\def\Sine#1%
       {{%
	\dimen 0 = #1 \r@dian
	\r@duce
	\ifdim\dimen0 = -90\r@dian \then
	   \dimen4 = -1\r@dian
	   \c@mputefalse
	\fi
	\ifdim\dimen0 = 90\r@dian \then
	   \dimen4 = 1\r@dian
	   \c@mputefalse
	\fi
	\ifdim\dimen0 = 0\r@dian \then
	   \dimen4 = 0\r@dian
	   \c@mputefalse
	\fi
	\ifc@mpute \then
		\divide\dimen0 by 180
		\dimen0=3.141592654\dimen0
		\dimen 2 = 3.1415926535897963\r@dian 
		\divide\dimen 2 by 2 
		\Mess@ge {Sin: calculating Sin of \nodimen 0}%
		\count 0 = 1 
		\dimen 2 = 1 \r@dian 
		\dimen 4 = 0 \r@dian 
		\loop
			\ifnum	\dimen 2 = 0 
			\then	\stillc@nvergingfalse 
			\else	\stillc@nvergingtrue
			\fi
			\ifstillc@nverging 
			\then	\term {\count 0} {\dimen 0} {\dimen 2}%
				\advance \count 0 by 2
				\count 2 = \count 0
				\divide \count 2 by 2
				\ifodd	\count 2 
				\then	\advance \dimen 4 by \dimen 2
				\else	\advance \dimen 4 by -\dimen 2
				\fi
		\repeat
	\fi		
			\xdef \sine {\nodimen 4}%
       }}

\def\Cosine#1{\ifx\sine\UnDefined\edef\Savesine{\relax}\else
		             \edef\Savesine{\sine}\fi
	{\dimen0=#1\r@dian\advance\dimen0 by 90\r@dian
	 \Sine{\nodimen 0}
	 \xdef\cosine{\sine}
	 \xdef\sine{\Savesine}}}	      

\def\psdraft{
	\def\@psdraft{0}
}
\def\psfull{
	\def\@psdraft{100}
}

\psfull

\newif\if@scalefirst
\def\psscalefirst{\@scalefirsttrue}
\def\psrotatefirst{\@scalefirstfalse}
\psrotatefirst

\newif\if@draftbox
\def\psnodraftbox{
	\@draftboxfalse
}
\def\psdraftbox{
	\@draftboxtrue
}
\@draftboxtrue

\newif\if@prologfile
\newif\if@postlogfile
\def\pssilent{
	\@noisyfalse
}
\def\psnoisy{
	\@noisytrue
}
\psnoisy
\newif\if@bbllx
\newif\if@bblly
\newif\if@bburx
\newif\if@bbury
\newif\if@height
\newif\if@width
\newif\if@rheight
\newif\if@rwidth
\newif\if@angle
\newif\if@clip
\newif\if@verbose
\def\@p@@sclip#1{\@cliptrue}
\newif\if@decmpr
\def\@p@@sfigure#1{\def\@p@sfile{null}\def\@p@sbbfile{null}\@decmprfalse
   \openin1=\ps@predir#1
   \ifeof1
	\closein1
	\get@dir{#1}
	\ifx\ps@founddir\leer
		\openin1=\ps@predir#1.bb
		\ifeof1
			\closein1
			\get@dir{#1.bb}
			\ifx\ps@founddir\leer
				\ps@typeout{Can't find #1 in \figurepath}
			\else
				\@decmprtrue
				\def\@p@sfile{\ps@founddir\ps@dir#1}
				\def\@p@sbbfile{\ps@founddir\ps@dir#1.bb}
			\fi
		\else
			\closein1
			\@decmprtrue
			\def\@p@sfile{#1}
			\def\@p@sbbfile{#1.bb}
		\fi
	\else
		\def\@p@sfile{\ps@founddir\ps@dir#1}
		\def\@p@sbbfile{\ps@founddir\ps@dir#1}
	\fi
   \else
	\closein1
	\def\@p@sfile{#1}
	\def\@p@sbbfile{#1}
   \fi
}
\def\@p@@sfile#1{\@p@@sfigure{#1}}
\def\@p@@sbbllx#1{
		\@bbllxtrue
		\dimen100=#1
		\edef\@p@sbbllx{\number\dimen100}
}
\def\@p@@sbblly#1{
		\@bbllytrue
		\dimen100=#1
		\edef\@p@sbblly{\number\dimen100}
}
\def\@p@@sbburx#1{
		\@bburxtrue
		\dimen100=#1
		\edef\@p@sbburx{\number\dimen100}
}
\def\@p@@sbbury#1{
		\@bburytrue
		\dimen100=#1
		\edef\@p@sbbury{\number\dimen100}
}
\def\@p@@sheight#1{
		\@heighttrue
		\dimen100=#1
   		\edef\@p@sheight{\number\dimen100}
}
\def\@p@@swidth#1{
		\@widthtrue
		\dimen100=#1
		\edef\@p@swidth{\number\dimen100}
}
\def\@p@@srheight#1{
		\@rheighttrue
		\dimen100=#1
		\edef\@p@srheight{\number\dimen100}
}
\def\@p@@srwidth#1{
		\@rwidthtrue
		\dimen100=#1
		\edef\@p@srwidth{\number\dimen100}
}
\def\@p@@sangle#1{
		\@angletrue
		\edef\@p@sangle{#1} 
}
\def\@p@@ssilent#1{ 
		\@verbosefalse
}
\def\@p@@sprolog#1{\@prologfiletrue\def\@prologfileval{#1}}
\def\@p@@spostlog#1{\@postlogfiletrue\def\@postlogfileval{#1}}
\def\@cs@name#1{\csname #1\endcsname}
\def\@setparms#1=#2,{\@cs@name{@p@@s#1}{#2}}
\def\ps@init@parms{
		\@bbllxfalse \@bbllyfalse
		\@bburxfalse \@bburyfalse
		\@heightfalse \@widthfalse
		\@rheightfalse \@rwidthfalse
		\def\@p@sbbllx{}\def\@p@sbblly{}
		\def\@p@sbburx{}\def\@p@sbbury{}
		\def\@p@sheight{}\def\@p@swidth{}
		\def\@p@srheight{}\def\@p@srwidth{}
		\def\@p@sangle{0}
		\def\@p@sfile{} \def\@p@sbbfile{}
		\def\@p@scost{10}
		\def\@sc{}
		\@prologfilefalse
		\@postlogfilefalse
		\@clipfalse
		\if@noisy
			\@verbosetrue
		\else
			\@verbosefalse
		\fi
}
\def\parse@ps@parms#1{
	 	\@psdo\@psfiga:=#1\do
		   {\expandafter\@setparms\@psfiga,}}
\newif\ifno@bb
\def\bb@missing{
	\if@verbose{
		\ps@typeout{psfig: searching \@p@sbbfile \space  for bounding box}
	}\fi
	\no@bbtrue
	\epsf@getbb{\@p@sbbfile}
        \ifno@bb \else \bb@cull\epsf@llx\epsf@lly\epsf@urx\epsf@ury\fi
}	
\def\bb@cull#1#2#3#4{
	\dimen100=#1 bp\edef\@p@sbbllx{\number\dimen100}
	\dimen100=#2 bp\edef\@p@sbblly{\number\dimen100}
	\dimen100=#3 bp\edef\@p@sbburx{\number\dimen100}
	\dimen100=#4 bp\edef\@p@sbbury{\number\dimen100}
	\no@bbfalse
}
\newdimen\p@intvaluex
\newdimen\p@intvaluey
\def\rotate@#1#2{{\dimen0=#1 sp\dimen1=#2 sp
		  \global\p@intvaluex=\cosine\dimen0
		  \dimen3=\sine\dimen1
		  \global\advance\p@intvaluex by -\dimen3
		  \global\p@intvaluey=\sine\dimen0
		  \dimen3=\cosine\dimen1
		  \global\advance\p@intvaluey by \dimen3
		  }}
\def\compute@bb{
		\no@bbfalse
		\if@bbllx \else \no@bbtrue \fi
		\if@bblly \else \no@bbtrue \fi
		\if@bburx \else \no@bbtrue \fi
		\if@bbury \else \no@bbtrue \fi
		\ifno@bb \bb@missing \fi
		\ifno@bb \ps@typeout{FATAL ERROR: no bb supplied or found}
			\no-bb-error
		\fi
		\count203=\@p@sbburx
		\count204=\@p@sbbury
		\advance\count203 by -\@p@sbbllx
		\advance\count204 by -\@p@sbblly
		\edef\ps@bbw{\number\count203}
		\edef\ps@bbh{\number\count204}
		\if@angle 
			\Sine{\@p@sangle}\Cosine{\@p@sangle}
	        	{\dimen100=\maxdimen\xdef\r@p@sbbllx{\number\dimen100}
					    \xdef\r@p@sbblly{\number\dimen100}
			                    \xdef\r@p@sbburx{-\number\dimen100}
					    \xdef\r@p@sbbury{-\number\dimen100}}
                        \def\minmaxtest{
			   \ifnum\number\p@intvaluex<\r@p@sbbllx
			      \xdef\r@p@sbbllx{\number\p@intvaluex}\fi
			   \ifnum\number\p@intvaluex>\r@p@sbburx
			      \xdef\r@p@sbburx{\number\p@intvaluex}\fi
			   \ifnum\number\p@intvaluey<\r@p@sbblly
			      \xdef\r@p@sbblly{\number\p@intvaluey}\fi
			   \ifnum\number\p@intvaluey>\r@p@sbbury
			      \xdef\r@p@sbbury{\number\p@intvaluey}\fi
			   }
			\rotate@{\@p@sbbllx}{\@p@sbblly}
			\minmaxtest
			\rotate@{\@p@sbbllx}{\@p@sbbury}
			\minmaxtest
			\rotate@{\@p@sbburx}{\@p@sbblly}
			\minmaxtest
			\rotate@{\@p@sbburx}{\@p@sbbury}
			\minmaxtest
			\edef\@p@sbbllx{\r@p@sbbllx}\edef\@p@sbblly{\r@p@sbblly}
			\edef\@p@sbburx{\r@p@sbburx}\edef\@p@sbbury{\r@p@sbbury}
		\fi
		\count203=\@p@sbburx
		\count204=\@p@sbbury
		\advance\count203 by -\@p@sbbllx
		\advance\count204 by -\@p@sbblly
		\edef\@bbw{\number\count203}
		\edef\@bbh{\number\count204}
}
\def\in@hundreds#1#2#3{\count240=#2 \count241=#3
		     \count100=\count240	
		     \divide\count100 by \count241
		     \count101=\count100
		     \multiply\count101 by \count241
		     \advance\count240 by -\count101
		     \multiply\count240 by 10
		     \count101=\count240	
		     \divide\count101 by \count241
		     \count102=\count101
		     \multiply\count102 by \count241
		     \advance\count240 by -\count102
		     \multiply\count240 by 10
		     \count102=\count240	
		     \divide\count102 by \count241
		     \count200=#1\count205=0
		     \count201=\count200
			\multiply\count201 by \count100
		 	\advance\count205 by \count201
		     \count201=\count200
			\divide\count201 by 10
			\multiply\count201 by \count101
			\advance\count205 by \count201
		     \count201=\count200
			\divide\count201 by 100
			\multiply\count201 by \count102
			\advance\count205 by \count201
		     \edef\@result{\number\count205}
}
\def\compute@wfromh{
		\in@hundreds{\@p@sheight}{\@bbw}{\@bbh}
		\edef\@p@swidth{\@result}
}
\def\compute@hfromw{
	        \in@hundreds{\@p@swidth}{\@bbh}{\@bbw}
		\edef\@p@sheight{\@result}
}
\def\compute@handw{
		\if@height 
			\if@width
			\else
				\compute@wfromh
			\fi
		\else 
			\if@width
				\compute@hfromw
			\else
				\edef\@p@sheight{\@bbh}
				\edef\@p@swidth{\@bbw}
			\fi
		\fi
}
\def\compute@resv{
		\if@rheight \else \edef\@p@srheight{\@p@sheight} \fi
		\if@rwidth \else \edef\@p@srwidth{\@p@swidth} \fi
}
\def\compute@sizes{
	\compute@bb
	\if@scalefirst\if@angle
	\if@width
	   \in@hundreds{\@p@swidth}{\@bbw}{\ps@bbw}
	   \edef\@p@swidth{\@result}
	\fi
	\if@height
	   \in@hundreds{\@p@sheight}{\@bbh}{\ps@bbh}
	   \edef\@p@sheight{\@result}
	\fi
	\fi\fi
	\compute@handw
	\compute@resv}
\def\OzTeXSpecials{
	\special{empty.ps /@isp {true} def}
	\special{empty.ps \@p@swidth \space \@p@sheight \space
			\@p@sbbllx \space \@p@sbblly \space
			\@p@sbburx \space \@p@sbbury \space
			startTexFig \space }
	\if@clip{
		\if@verbose{
			\ps@typeout{(clip)}
		}\fi
		\special{empty.ps doclip \space }
	}\fi
	\if@angle{
		\if@verbose{
			\ps@typeout{(rotate)}
		}\fi
		\special {empty.ps \@p@sangle \space rotate \space} 
	}\fi
	\if@prologfile
	    \special{\@prologfileval \space } \fi
	\if@decmpr{
		\if@verbose{
			\ps@typeout{psfig: Compression not available
			in OzTeX version \space }
		}\fi
	}\else{
		\if@verbose{
			\ps@typeout{psfig: including \@p@sfile \space }
		}\fi
		\special{epsf=\ps@predir\@p@sfile \space }
	}\fi
	\if@postlogfile
	    \special{\@postlogfileval \space } \fi
	\special{empty.ps /@isp {false} def}
}
\def\DvipsSpecials{
	\special{ps::[begin] 	\@p@swidth \space \@p@sheight \space
			\@p@sbbllx \space \@p@sbblly \space
			\@p@sbburx \space \@p@sbbury \space
			startTexFig \space }
	\if@clip{
		\if@verbose{
			\ps@typeout{(clip)}
		}\fi
		\special{ps:: doclip \space }
	}\fi
	\if@angle
		\if@verbose{
			\ps@typeout{(clip)}
		}\fi
		\special {ps:: \@p@sangle \space rotate \space} 
	\fi
	\if@prologfile
	    \special{ps: plotfile \@prologfileval \space } \fi
	\if@decmpr{
		\if@verbose{
			\ps@typeout{psfig: including \@p@sfile.Z \space }
		}\fi
		\special{ps: plotfile "`zcat \@p@sfile.Z" \space }
	}\else{
		\if@verbose{
			\ps@typeout{psfig: including \@p@sfile \space }
		}\fi
		\special{ps: plotfile \@p@sfile \space }
	}\fi
	\if@postlogfile
	    \special{ps: plotfile \@postlogfileval \space } \fi
	\special{ps::[end] endTexFig \space }
}
\def\psfig#1{\vbox {
	\ps@init@parms
	\parse@ps@parms{#1}
	\compute@sizes
	\ifnum\@p@scost<\@psdraft{
		\PsfigSpecials 
		\vbox to \@p@srheight sp{
			\hbox to \@p@srwidth sp{
				\hss
			}
		\vss
		}
	}\else{
		\if@draftbox{		
			\hbox{\fbox{\vbox to \@p@srheight sp{
			\vss
			\hbox to \@p@srwidth sp{ \hss 
			 \hss }
			\vss
			}}}
		}\else{
			\vbox to \@p@srheight sp{
			\vss
			\hbox to \@p@srwidth sp{\hss}
			\vss
			}
		}\fi

	}\fi
}}
\psfigRestoreAt
\setDriver
\let\@=\LaTeXAtSign

\ifoldfss
  \newcommand{\rmn}[1] {{\rm #1}}
  \newcommand{\itl}[1] {{\it #1}}
  \newcommand{\bld}[1] {{\bf #1}}
  \ifCUPmtlplainloaded \else
    \NewTextAlphabet{textbfit} {cmbxti10} {}
    \NewTextAlphabet{textbfss} {cmssbx10} {}
    \NewMathAlphabet{mathbfit} {cmbxti10} {} 
    \NewMathAlphabet{mathbfss} {cmssbx10} {} 
  \fi
  \ifAMStwofonts
    \ifCUPmtlplainloaded \else
      \NewSymbolFont{upmath} {eurm10}
      \NewSymbolFont{AMSa} {msam10}
      \NewMathSymbol{\upi}     {0}{upmath}{19}
      \NewMathSymbol{\umu}     {0}{upmath}{16}
      \NewMathSymbol{\upartial}{0}{upmath}{40}
      \NewMathSymbol{\leqslant}{3}{AMSa}{36}
      \NewMathSymbol{\geqslant}{3}{AMSa}{3E}
      \let\oldle=\le     \let\oldleq=\leq
      \let\oldge=\ge     \let\oldgeq=\geq
      \let\leq=\leqslant \let\le=\leqslant
      \let\geq=\geqslant \let\ge=\geqslant
    \fi
  \fi
\fi 

\ifnfssone
  \newmathalphabet{\mathit}
  \addtoversion{normal}{\mathit}{cmr}{m}{it}
  \addtoversion{bold}{\mathit}{cmr}{bx}{it}
  \newcommand{\rmn}[1] {\mathrm{#1}}
  \newcommand{\itl}[1] {\mathit{#1}}
  \newcommand{\bld}[1] {\mathbf{#1}}
  \def\textbfit{\protect\txtbfit}
  \def\textbfss{\protect\txtbfss}
  \long\def\txtbfit#1{{\fontfamily{cmr}\fontseries{bx}\fontshape{it}%
    \selectfont #1}}
  \long\def\txtbfss#1{{\fontfamily{cmss}\fontseries{bx}\fontshape{n}%
    \selectfont #1}}
  \newmathalphabet{\mathbfit} 
  \addtoversion{normal}{\mathbfit}{cmr}{bx}{it}
  \addtoversion{bold}{\mathbfit}{cmr}{bx}{it}
  \newmathalphabet{\mathbfss} 
  \addtoversion{normal}{\mathbfss}{cmss}{bx}{n}
  \addtoversion{bold}{\mathbfss}{cmss}{bx}{n}
  \ifAMStwofonts
    \ifCUPmtlplainloaded \else
      \UseAMStwoboldmath
      \makeatletter
      \new@mathgroup\upmath@group
      \define@mathgroup\mv@normal\upmath@group{eur}{m}{n}
      \define@mathgroup\mv@bold\upmath@group{eur}{b}{n}
      \edef\UPM{\hexnumber\upmath@group}
      \new@mathgroup\amsa@group
      \define@mathgroup\mv@normal\amsa@group{msa}{m}{n}
      \define@mathgroup\mv@bold\amsa@group{msa}{m}{n}
      \edef\AMSa{\hexnumber\amsa@group}
      \makeatother
      \mathchardef\upi="0\UPM19
      \mathchardef\umu="0\UPM16
      \mathchardef\upartial="0\UPM40
      \mathchardef\leqslant="3\AMSa36
      \mathchardef\geqslant="3\AMSa3E
      \let\oldle=\le     \let\oldleq=\leq
      \let\oldge=\ge     \let\oldgeq=\geq
      \let\leq=\leqslant \let\le=\leqslant
      \let\geq=\geqslant \let\ge=\geqslant
    \fi
  \fi
\fi 

\ifnfsstwo
  \newcommand{\rmn}[1] {\mathrm{#1}}
  \newcommand{\itl}[1] {\mathit{#1}}
  \newcommand{\bld}[1] {\mathbf{#1}}
  \def\textbfit{\protect\txtbfit}
  \def\textbfss{\protect\txtbfss}
  \long\def\txtbfit#1{{\fontfamily{cmr}\fontseries{bx}\fontshape{it}%
    \selectfont #1}}
  \long\def\txtbfss#1{{\fontfamily{cmss}\fontseries{bx}\fontshape{n}%
    \selectfont #1}}
  \DeclareMathAlphabet{\mathbfit}{OT1}{cmr}{bx}{it}
  \SetMathAlphabet\mathbfit{bold}{OT1}{cmr}{bx}{it}
  \DeclareMathAlphabet{\mathbfss}{OT1}{cmss}{bx}{n}
  \SetMathAlphabet\mathbfss{bold}{OT1}{cmss}{bx}{n}
  \ifAMStwofonts
    \ifCUPmtlplainloaded \else
      \DeclareSymbolFont{UPM}{U}{eur}{m}{n}
      \SetSymbolFont{UPM}{bold}{U}{eur}{b}{n}
      \DeclareSymbolFont{AMSa}{U}{msa}{m}{n}
      \DeclareMathSymbol{\upi}{0}{UPM}{"19}
      \DeclareMathSymbol{\umu}{0}{UPM}{"16}
      \DeclareMathSymbol{\upartial}{0}{UPM}{"40}
      \DeclareMathSymbol{\leqslant}{3}{AMSa}{"36}
      \DeclareMathSymbol{\geqslant}{3}{AMSa}{"3E}
      \let\oldle=\le     \let\oldleq=\leq
      \let\oldge=\ge     \let\oldgeq=\geq
      \let\leq=\leqslant \let\le=\leqslant
      \let\geq=\geqslant \let\ge=\geqslant
    \fi
  \fi
\fi 

\ifCUPmtlplainloaded \else
  \ifAMStwofonts \else 
    \def\upi{\pi}
    \def\umu{\mu}
    \def\upartial{\partial}
  \fi
\fi

\title[Microlensed GRB light curves]{Distortion of gamma-ray burst light 
curves by gravitational microlensing}
\author[L. L. R. Williams and R. A. M. J. Wijers]
       {L. L. R. Williams and R. A. M. J. Wijers \\
        Institute of Astronomy, Madingley Road, Cambridge, CB3 0HA\\
       Email: {\tt llrw@ast.cam.ac.uk}  and {\tt ramjw@ast.cam.ac.uk}\\}
\date{Submitted 3 September, 1996}
\pubyear{1996}

\begin{document}

\maketitle

\newcommand{\fmmm}[1]{\mbox{$#1$}}
\newcommand{\scnd}{\mbox{\fmmm{''}\hskip-0.3em .}}
\newcommand{\scnp}{\mbox{\fmmm{''}}}

\begin{abstract}
If at cosmological distances, a small fraction of gamma-ray bursts should be
multiply imaged by intervening galaxies or clusters, resulting in the 
appearance of two very similar bursts from the same location with a relative
time delay of hours to a year. We show
that microlensing by individual stars in the lensing galaxy can smear out
the light curves of the multiply imaged bursts on millisecond time scales.
Therefore, in deciding whether two bursts are similar enough to qualify as 
multiple images, one must look at time scales longer than a few tens of
milliseconds, since shorter time scales are possibly rendered dissimilar
by microlensing.
\end{abstract}


\section{Introduction}

After more than two decades of observations, Gamma-Ray Bursts (GRBs) remain
arguably the most puzzling phenomena in astronomy. In particular, their
distance scale is yet to be established to everyone's satisfaction, with
100\,kpc (Galactic corona) and $z=1$ ($\sim$ few Gpc) being the main 
opposing candidates (see Fishman 1995).
There are two major pieces of evidence that point to the cosmological
origin of the GRBs. First, evidence of redshift in faint gamma-ray bursts
has emerged, both from time dilation in faint bursts relative to bright
ones (Norris et al.\ 1994, 1995) and from the observation that the peak
energies of the spectra are lower for faint bursts (Malozzi et al.\
1995), each effect placing the faint bursts at a redshift of about 1.  Both
effects are only detectable in a statistical sense because the durations
and hardnesses of gamma-ray bursts range widely at any given brightness,
and subtleties of data analysis therefore play a role, leading others to
not detect the time dilation (Mitrofanov et al.\ 1996). Also, there could be
intrinsic correlations between the different properties of a gamma-ray
burst that mimic redshift effects, though they are distinguishable in
principle from the real thing, and in fact a cosmological origin agrees
better with the detected time dilation (see Wijers \& Paczy\'nski 1994).

Second, evidence has been found of correlations between bright gamma-ray
bursts and objects at their expected redshift (if cosmological) of a few
tenths.  Recently, Rood \& Struble (1996) and Marani et al.  (1996)
showed that the brighter of the GRBs
are correlated with Abell (ACO) clusters. Larson, McLean, and Becklin 
(1996) found an
excess of galaxies at K band near well-localized gamma-ray bursts.
The significance of these is not yet very great, but may
improve with time.

It has been pointed out that strong gravitational lensing, resulting in
a GRB being split into more than one macroimage, may provide the best means
to ascertain the cosmological origin of GRBs (Paczy\'nski 1986b). If a
multiply imaged lensing event is observed, then the source can be assumed 
to be at a
cosmological distance from us. How do we know if two bursts originating
from the same location, to within the errors, are multiply imaged and are
not due to a repeater or are simply two unrelated events? If the
observations are free of noise, and microlensing by stars is unimportant,
then the multiple images (macroimages) due to an intervening galaxy or
cluster should have identical spectra and light curves, to within a scale
factor equal to the relative image magnifications. The presence of noise
will degrade the light curves, thus macroimages can appear dissimilar. The
effect of noise and faintness of the macroimages on their potential
classification as `identical' was discussed by Wambsganss (1993) and Nowak
\& Grossman (1994). Here we consider microlensing by individual
stars in the lensing galaxy, and show that it can `smear' image light
curves on time scales of up to a few tens of milliseconds, making multiple
images of the same burst even less similar.

Should one expect to see multiply imaged GRBs if they are at cosmological
distances?
Mao (1992) and Grossman \& Nowak (1994) estimated that the waiting time for 
one lensed pair to show up in the BATSE catalogue can vary from one to well
over ten years, depending on cosmology and other assumptions, so it is
by no means certain that a lensed pair will be found with BATSE. 

Microlensing by stars is known to be important for some multiply imaged
QSOs, for example, Q2237+0305, the Einstein Cross (Irwin et al. 1989, Houde
\& Racine 1994), and probably H1413+117, the Clover Leaf (Arnould et al.
1993), PG1115+080 (Vanderriest et al. 1986) and possibly a few more QSOs.
Therefore it is expected that macroimages of GRBs will also be affected by
microlensing. For concreteness, we examine one particular case that can
hypothetically be observed. This case roughly corresponds to the image
parameters of a well known lens: Q2237+0305 (Wambsganss et al. 1990). 
We show what would have been observed if the source here were a GRB instead 
of a QSO.

Throughout this paper we assume a standard cosmology with $\Omega=1$, and\\
$H_0=75\,$km\,s$^{-1}$\,Mpc$^{-1}$.

\section{Examples of microlensed gamma-ray bursts}

\subsection{Microlensing calculations}

The gravitational potential of the lensing galaxy splits a single
background burst into multiple macroimages, whose arrival times can
differ by anywhere between a few hours to over a year, and whose 
angular separation will generally be less than several arcseconds. 
Because the galaxies are made up of stars, the 
lensing mass of the galaxy is grainy, and therefore the macroimages are 
also grainy, i.e. they consist of microimages that have a spatial as 
well as a temporal distribution. Given the poor angular resolution of
gamma-ray telescopes, not only the microimages, but also the macroimages 
cannot be precisely localized. On the other hand, the temporal resolution 
of some of the brighter GRBs is very good, of order milliseconds, 
and so the 
effects of microlensing are important here: the observed macroimages will 
appear as a sum of many staggered and scaled versions of the original burst.

\begin{figure}
\vbox{
\centerline{
\psfig{figure=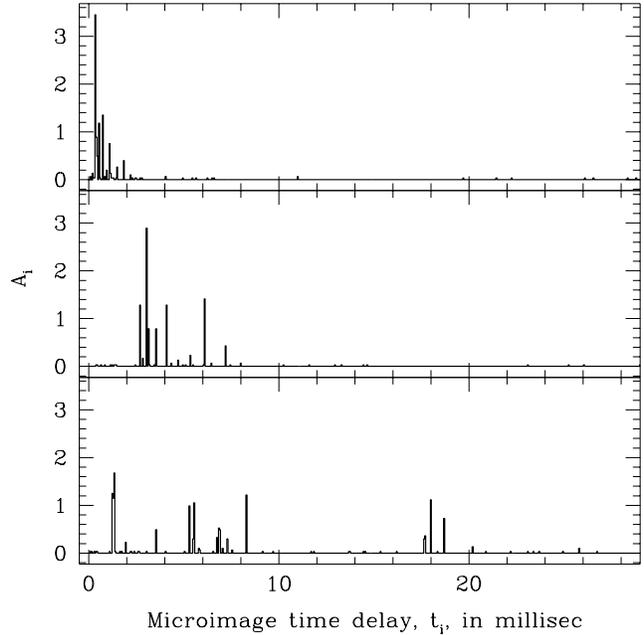,width=3.5in,angle=0}
}
\caption[]{Microlensing light curves of a single $\delta$-function burst.
The macrolensing parameters are $\kappa$=0.524, $\gamma=0.354$, with a
total mean magnification of $\bar A=9.86$. Horizontal axis is the
time delay in milliseconds, and vertical axis is the brightness of 
individual microimages, such that the total unmagnified brightness of the 
source is unity. Notice the diversity of microlensing profiles, and that
the total time width generally exceeds several milliseconds.
\label{mlightcurves}
}
}
\end{figure}

To quantify the above description, we first need to specify a macrolensing
model, and hence the properties of the macroimages. Given these, we can
obtain a temporal profile of a microlensed $\delta$-function burst, and 
later convolve it with an observed GRB light curve to produce an example 
of a possible microlensed GRB profile.

We start with a macrolensing model of a galaxy, given by a non-singular
isothermal sphere, with a core radius (See eq. [8.37] and section 8.1.5 
of Schneider et al. 1992 for details). Let us now generate a set of
macrolensing parameters reminiscent of the actual observed QSO lens,
Q2237+0305. In that case, five images are observed, four of similar 
magnification, and a demagnified central image. In our case, since the
lens model is assumed to be circularly symmetric (for simplicity), only
three images will be produced, two images of comparable magnification,
and a demagnified third one. Let the galaxy core radius be, $r_c$=0.36 kpc
($\theta_c=0.5^{\prime\prime}$), and its velocity dispersion, 
$\sigma_v$=185 km/s. The source and lens redshifts are $z_s=1.7$, and 
$z_l=0.04$, respectively. For a source impact parameter, 
$\theta_s=0.125^{\prime\prime}$, three images are formed. The two brighter 
macroimages have magnifications of 9.86 and 9.31, a separation of 
$1.72^{\prime\prime}$, and a time delay of 0.92 days. At the locations 
of these macroimages, the optical depth and shear ($\kappa$, $\gamma$) 
due to the galaxy are (0.524, 0.354), and (0.799, 0.385), respectively.
These parameters are required to generate the microimage distributions.

\begin{figure}
\vbox{
\centerline{
\psfig{figure=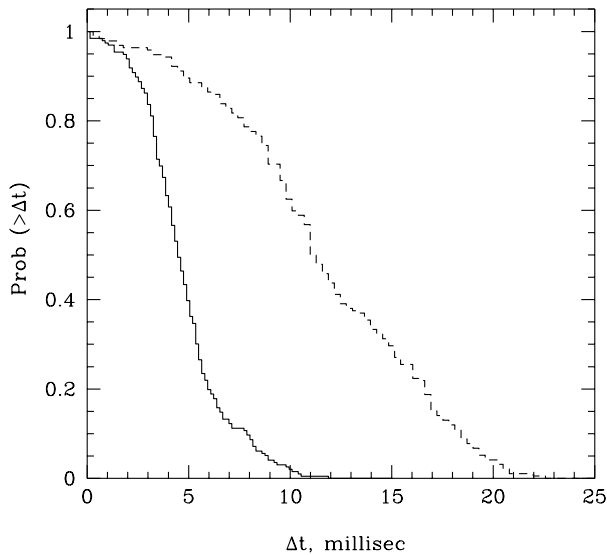,width=3.25in,angle=0}
}
\caption[]{Normalized cumulative histograms of time delay spread of 
microimages, $\Delta t$ in milliseconds, for a set macroimages
with $\kappa$=0.799, $\gamma=0.385$ (solid line), and  
$\kappa$=0.524, $\gamma=0.354$ (dashed line).
\label{histogram}
}
}
\end{figure}

Given a set of $\kappa$ and $\gamma$ for the first of the two macroimages, 
we can now numerically simulate the action of microlensing on a 
$\delta$-function burst of radiation from a small source. The patch of 
the lensing galaxy where the macroimage is formed is represented by a 
two-dimensional random distribution of stars. The mass function of stars is
taken from Scalo (1986); it has an average mass of $\bar M=0.386 M_\odot$, 
and an upper and lower mass cutoffs of 63$M_\odot$ and 0.087$M_\odot$ 
respectively.
A ray-tracing method implemented using a hierarchical tree code is used 
to find the microimages of a small source. Once the locations of the 
microimages are found, the geometrical and gravitational parts of the 
time delay are calculated for each microimage separately. 
Figure~\ref{mlightcurves} shows three representative examples of the 
resulting time series. Note that the three light curves look very different
from each other, which is expected since microlensing at even moderate 
optical depths is a highly non-linear process. A good way to parameterize
a given light curve is to calculate its brightness-weighted rms spread
$\Delta t$,
$$\Delta t=\sqrt{{{\sum_{i=1}^N(t_i-{\bar t})^2~A_i}\over{\sum_{i=1}^N~A_i}}},
{\hskip 0.1in}{\rm where}{\hskip 0.15in}
{\bar t}={{\sum_{i=1}^N~t_i A_i}\over{\sum_{i=1}^N~A_i}},$$
$N$ is the total number of microimages in the lightcurve;
$t_i$'s and $A_i$'s are their individual arrival times and 
magnifications, respectively.
It is evident from Figure~\ref{mlightcurves} that the values 
of $\Delta t$ can vary appreciably, even for a fixed set of ($\kappa$, 
$\gamma$). Figure~\ref{histogram} is a normalized cumulative histogram
of $\Delta t$ values for many independent macroimages having the same
$\kappa$, $\gamma$ values: the solid line is for (0.799, 0.385), and the 
dashed one for (0.524, 0.354). The median $\Delta t$ is 5--10
milliseconds, and is therefore relevant for some GRBs. 

   \subsection{Application to real GRB light curves}

\begin{figure*}
\begin{center}
   \epsfbox{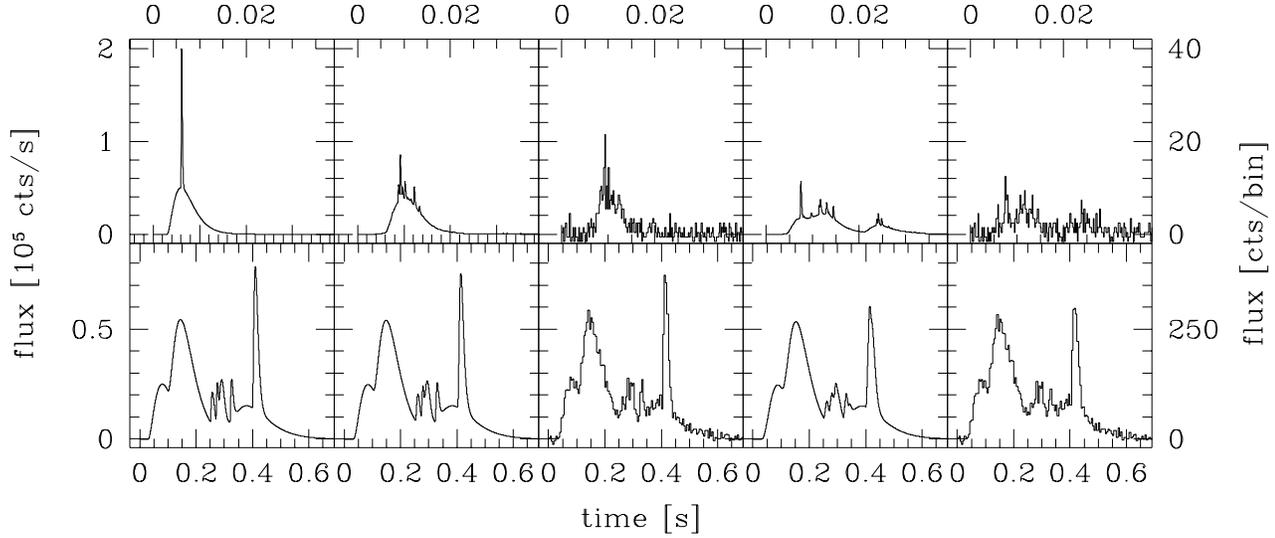}
\end{center}
\caption[]{A short (top row) and long (bottom row) synthetic gamma-ray 
burst affected by microlensing. The leftmost panel shows the ideal burst.
The next two panels show how it is changed by folding the ideal GRB with
the top microlensing pattern of Figure~\ref{mlightcurves}, before and
after binning and adding realistic noise. The next panels are the same
for folding with the bottom microlensing pattern of 
Figure~\ref{mlightcurves}.
\label{convolution}
}
\end{figure*}

The magnitude and detectability of microlensing distortions depend both on
the temporal structure and signal strength of the original burst. To 
illustrate this, we convolve two of the three microlensing patterns of
Figure~\ref{mlightcurves} with two rather different bursts. For this 
purpose we constructed two time series that after binning and addition
of appropriate amounts of noise passably resemble BATSE bursts 1B\,910711
and 1B\,920218B (the latter shortened by a factor 4 in time). The former
was one of the shortest ever seen and contained a sub-millisecond spike
(Bhat et al.\ 1992); the latter is more typical.
Figure~\ref{convolution} shows a convolution of
microlensing time profiles with the two simulated GRB light curves. As can
be seen, the short burst can be distorted beyond recognition, so
it would take some courage to advance the view that they are a
lensed pair (the paucity of such short bursts would help, of course).
The second case is more promising, for its high signal
to noise ratio allows one to both recognize the overall similarity at most
time scales and the differences at the shortest ones (especially the width
of the last spike). From a visual inspection of the first BATSE catalogue
(Fishman et~al.\ 1994) it appears that some 20 bursts have visible structure
at time scales of 10 ms or shorter, out of 40--50 in which the plot allows
one to identify such structure, so the fraction of bursts in which signs
of microlensing could be detected if they were a member of a lensed pair
may be as high as 50\%.

\section{Time width of the microimage distribution}\label{analytical}

It is interesting to note that the temporal extent of microimages
(Figure~\ref{mlightcurves}) is much larger than the time delay between two
images created by a single isolated star. In the latter case, time delay is
approximately $8GM/c^3$, which is about 0.04 ms for a solar mass
star-lens if the images are of comparable brightness. This difference
between the time width of microimages in a galaxy and those due to an
isolated star occurs because a collection of stars, as in a galaxy, act
together to produce a wider $\Delta t$ than a single isolated star. 

Before
we proceed to derive a rough analytic expression for the total time spread 
of a microlensed burst, we note that our analysis is related to that of 
Katz, Balbus \& Paczy\'nski (1986). However, there is a difference between 
the two approaches that makes Katz et al. result not directly applicable 
to our case. Just like we do below,
Katz et al. look at the distribution of microimage magnifications. They
pose a question, ``What is the size of the region in the lens plane that
catches a given percentage of the macroimage flux?''. Below, we answer 
a somewhat different question, ``At what distance from the source does the 
magnification of individual microimages become a certain fraction of the
macroimage flux?''. 

Imagine a two-dimensional star field of microlenses of optical depth
$\kappa$, and let all stars have the same mass of one mass unit. Consider a
background source whose unperturbed position coincides with the centre of
the coordinate system. Roughly, there will be at least $N+1$ images formed: 
a primary image, and one image associated with every star-microlens in the
vicinity of the source. In other words, the macroimage will consist of a
`cloud' of microimages surrounding the unperturbed source position. (For a
pictorial representation of a microlensing `cloud' see Paczy\'nski 1986a, 
and Wambsganss 1990). 
The size of the cloud is directly related to the geometrical time delay of 
the microimages. Since the gravitational and geometrical parts of the time 
delay are of the same order, one simply needs to know the size of the cloud
in order to estimate the total time delay of microimages.
We will now calculate the radius of this cloud.

Consider a microlens on the $x$-axis, a distance $L$ from the origin, where
$L$ is large compared to the Einstein ring radius of a microlens.  An image
due to this microlens will be located at $x=L+x^\prime$, where $x^\prime$
is small.  There are two contributions to the deflection angle in this
case: the microlens itself, and a sheet of microlenses interior to $L$. On
average, the mass of this sheet is $\kappa L^2$. The total deflection angle
is then given by $\alpha_x=\kappa L^2/x + 1/x^\prime$, and $\alpha_y=0$. The
magnification of the microimage,  $A^{-1}=1-\gamma_\star^2$,
is due to shear only%
\footnote{Note
that this shear, $\gamma_\star$, is different from the external shear,
$\gamma$. The latter is due to the overall potential of the galaxy, while
the former is due to the matter distribution close to the macroimage.}%
, since by assumption there is no continuously distributed matter.
The shear is given by $\gamma_\star=\kappa-{x^\prime}^{-2}\sim
{x^\prime}^{-2}$. Solving the lensing equation, $0=x-\kappa
L^2/x-1/x^\prime$, for the location of the images, one finds that
$1/x^\prime\sim (1-\kappa)L$. Therefore, we arrive at the following relation
between the source--microlens separation and the magnification of the
corresponding microimage, $L^4\sim A^{-1}(1-\kappa)^{-4}$. 

So far we have
assumed external shear to be 0, therefore the distribution of 
microimages is circularly symmetric with the diameter proportional to
$(1-\kappa)^{-1}$. If external shear is introduced, the overall shape
of the microimage cloud will get elongated, with the major and minor axes
proportional to $(1-\kappa-\gamma)^{-1}$ and $(1-\kappa+\gamma)^{-1}$ 
respectively\footnote{This is valid if $\kappa<1$. If $\kappa>1$, the major
and minor axes are proportional to $(1-\kappa+\gamma)^{-1}$ and 
$(1-\kappa-\gamma)^{-1}$.}. Since it is the longer of the axes that
determines the total width of microimage distribution, the final expression 
is, $L^4\sim A^{-1}(1-\kappa-\gamma)^{-4}$.

The total width of the arrival times of various microimages is roughly the
time delay of the outermost microimage of sufficiently non-small $A$. In
our dimensionless units, a single isolated microlens will produce a time
delay of 1 between microimages of comparable brightness.  By comparison, a
sheet of microlenses will have a width of time delays of $L^2\sim
A^{-1/2}(1-\kappa-\gamma)^{-2}$. For example, if images as faint as 0.01 
times the source brightness are considered, i.e. $A=0.01$, and 
$\kappa=0.524$, and $\gamma=0.354$ (as before) the total temporal extent
of microimages is $671$ dimensionless units, or, accounting for the
$\bar M$ and $(1+z_l)$ time dilation factor, about 10.8 milliseconds.

\begin{figure}
\vbox{
\centerline{
\psfig{figure=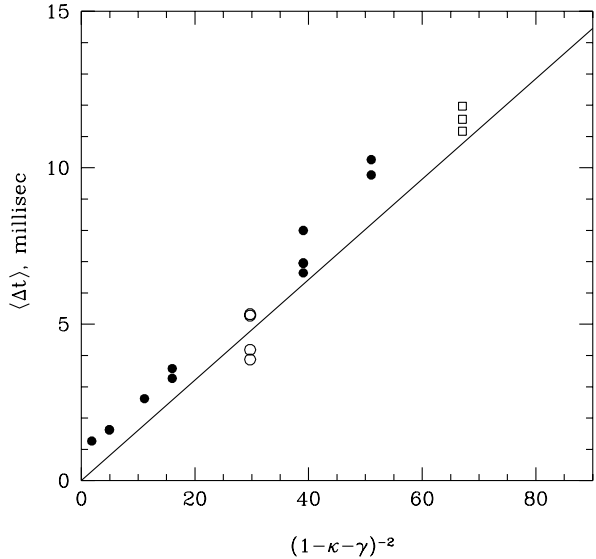,width=3.25in,angle=0}
}
\caption[]{Analytically-derived (straight line, see Section~\ref{analytical})
and numerically calculated (points) relation between the rms spread of 
microlensing delays and the macrolensing quantity, $(1-\kappa-\gamma)^{-2}$.
Empty circles and squares correspond to the ($\kappa$, $\gamma$) cases
represented by solid and dashed histograms, respectively, in 
Figure~\ref{histogram}. The straight line is, $\langle\Delta t\rangle=
(1+z_l)~(8GM/c^3)~{\bar M}~A^{-1/2}~(1-\kappa-\gamma)^{-2}$, where $z_l=0.04$
(reminiscent of the 2237+0305 quadruple QSO lens), $A=0.01$ (see text), 
and ${\bar M}=0.386M_\odot$.
\label{correlation}
}
}
\end{figure}

The straight line in Figure~\ref{correlation} is the analytical derivation
presented above. The points are the results of our numerical simulations 
using the ray-tracing code. The y-coordinate of each point, 
$\langle\Delta t\rangle$,
is the average of $\Delta t$, as defined at the end of Section 2.1, over 
many realizations of the source position. A range of $\kappa$, $\gamma$ 
parameters were used to test the analytical formula. The fit seems to be 
good considering the simplicity of our derivation.

\section{GRB observables and microlensing}

In general, all macroimages of a multiply images GRB will be affected 
by microlensing to some
degree, since the quantity $(1-\kappa-\gamma)^{-2}$ of macroimages is 
usually greater than a few. However what one really wants to know, when 
faced with a pair of candidate macrolensing events, is how much that 
particular pair has been affected. We will now show that there is 
insufficient information in the two gamma-ray bursts to estimate this 
independently.

A macroimage is strongly affected by microlensing if the temporal extent
of microimages is sufficiently large.  The parameter that determines the 
temporal spread of the microimages is $(1-\kappa-\gamma)^{-2}$.
Therefore, one would want to know this quantity for each of the images.

\begin{figure}
\vbox{
\centerline{
\psfig{figure=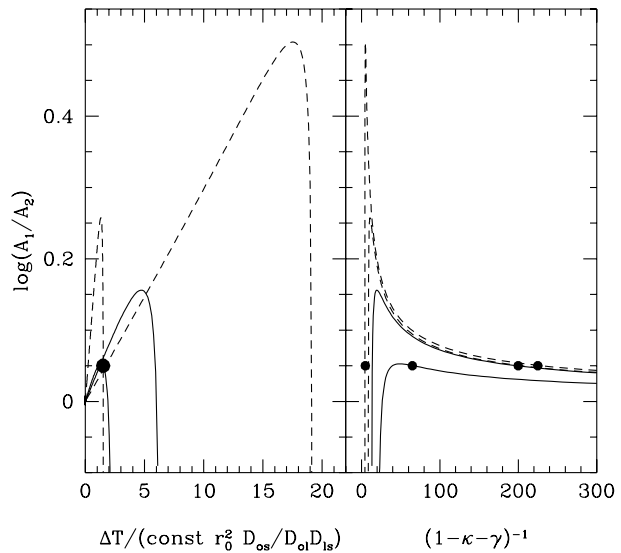,width=3.25in,angle=0}
}
\caption[]{Relative macroimage magnifications, log$(A_1/A_2)$, vs. relative
time delay, $\Delta T$, and $(1-\kappa-\gamma)^{-1}$ of the primary image.
Four macrolens models are considered: two non-singular isothermal spheres,
and two de Vaucouleurs profiles, each with a different, but realistic,
central surface mass densities. Notice that a single set of log$(A_1/A_2)$
and $\Delta T$ (large dot on the left panel) can correspond to four very
different $(1-\kappa-\gamma)^{-1}$ (four dots on the right panel). This 
example demonstrates that GRB observables are insufficient to determine 
the extent to which macroimages have been affected by microlensing. 
\label{macromodel} }
}
\end{figure}

Given a pair of suspected macroimages, two parameters can be observed: the
ratio of the total fluxes of the GRBs, which is roughly the magnification
ratio of the two macroimages, $A_1/A_2$, and their relative arrival time,
$\Delta T$.
The filled dot on the left panel of
Figure~\ref{macromodel} is an example of such a set of observables. The
lines correspond to log$(A_1/A_2)$ vs. dimensionless $\Delta T$ 
of the brighter
of the two images produced by a non-singular isothermal (solid lines), and
a de Vaucouleurs (dashed lines) lens model. The two lines of each model
have different central mass densities; since the identity of the galaxy
lens will not be known in such cases, its central mass density will be
unknown to within a factor of 2--3. The four dots on the right panel of
Figure~\ref{macromodel} show the corresponding $(1-\kappa-\gamma)^{-1}$ 
values of the primary image, in these four models. It is seen that a set 
of $A_1/A_2$ and $\Delta T$ does not constrain the $(1-\kappa-\gamma)^{-1}$
of the macroimage.

To make things worse, there is uncertainty in the values of $A_1/A_2$ and
$\Delta T$. The ratio of magnifications can be affected by microlensing,
realistically, by up to a factor of 2, or 0.3 in the log. The relative time
delay is uncertain because it is proportional to
$(D_{os}/D_{ol}D_{ls})r_0^2$, where $D$'s are the angular diameter
distances between the source, lens and observer, and $r_0$ is the length
scale appropriate to the macrolens model: core radius for the isothermal
model, and effective radius for the de Vaucouleurs model. For realistic
ranges of GRB source redshifts, 0.8--1.2, and lens redshift, 0.2--0.4, there
is a 30--50\% uncertainty in the cosmological parameter value; and a
factor of at least 2 uncertainty in the core/effective radius. Thus there
is a factor of at least 4--5 uncertainty in the value of $\Delta T$, and 
our lack of knowledge about the particulars of the lensing galaxy makes
the determination of $(1-\kappa-\gamma)^{-1}$ virtually impossible.

Finally, $(1-\kappa-\gamma)^{-1}$ is related to the {\it average} of  
$\Delta t$'s; in any particular case, $\Delta t$ can assume a range 
of values, even if $(1-\kappa-\gamma)^{-1}$ is known precisely 
(see Figures~\ref{mlightcurves} and \ref{histogram}).


\section{Summary}

It has already been pointed out (Wambsganss 1993) that macrolensed copies of
a single GRB need not look alike due to noise, and faintness. In this
paper we discuss another reason: microlensing due to stars in the lensing
galaxy. Most of the models of GRBs predict the `emitting region' to be much 
smaller than the Einstein ring radii of even the smallest lenses. The former
is of the order of $10^{13}$cm, while the latter is roughly $10^{14}$cm for 
a $10^{-5}M_\odot$ compact object. The microlensing light curve of a short 
flash originating from a compact region
is a series of spikes; the spikes are due to individual microimages, each
with a different arrival time. The width of their time distribution is a
function of the optical depth and shear of the galaxy at the location of 
the macroimages, and is 5--10 milliseconds for typical values of 
($\kappa$,$\gamma$) of the lensing galaxy, and a typical stellar mass 
function, with an average mass of a few tenths of $M_\odot$. Therefore GRB 
light curves are expected to be affected on these time scales. Similarly, 
if GRB spectra are varying on these time scales, they too will be affected.

\section*{Acknowledgments}

We would like to thank Andrew Lyne for a lively discussion that inspired
this work, and Martin Rees for useful discussions. We acknowledge the 
support of PPARC fellowships at the Institute of Astronomy, Cambridge.

\end{document}